\documentclass{aa}  
\usepackage{graphicx}
\usepackage{subcaption} 
\usepackage[varg]{txfonts}
\usepackage[utf8]{inputenc}
\usepackage{textgreek}
\usepackage{natbib}
\bibpunct{(}{)}{;}{a}{}{,} 
\usepackage{balance}
\usepackage{multirow}
\usepackage{hyperref}
\usepackage{xcolor}
\begin{document} 
\title{Deepest ever photographed Geminid with small but not negligible terminal mass}
\author{Pavel Spurn\'y \and Ji\v{r}\'i Borovi\v{c}ka} 

   \institute{Astronomical Institute of the Czech Academy of Sciences, Fri\v{c}ova 298, 25165 Ond\v{r}ejov, Czech Republic \\
              \email{pavel.spurny@asu.cas.cz}}

\date{Received date 22 January 2026 / Accepted date 18 February 2026} 
\abstract{
We report an instrumental observation of the very exceptional Geminid fireball which was observed in scope of the Czech part of the European Fireball Network (EN) on 13 December 2012 at 4$^h$12$^m$59.4$^s$ UT. The uniqueness of this Geminid fireball consists of the record depth of its penetration in the atmosphere (to the height of 32.5 km) and in the fact that most likely a very small fraction of its initial mass survived severe deceleration in the atmosphere and landed on the ground. Such deeply penetrating Geminid with so precise and reliable data has not yet been observed. From a comparison with a large number of Geminids observed by the European Fireball Network and all brightest Geminids from the Prairie Fireball Network in USA and the Canadian MORP Network, we have shown that for Geminids with an entry mass greater than approximately 10 grams, the terminal altitude limit does not decrease further as it does for smaller Geminids, but remains constant at around 38 km. In this comparison, we have shown that there is only one exception, and that is the Geminid presented here. This one penetrated nearly 6 km deeper with very low terminal speed for Geminids. During the atmospheric flight this Geminid meteoroid slowed down from its original speed of 35.75 km s$^{-1}$ to 6.8 km s$^{-1}$. This small meteoroid with initial mass of only 0.25 kg is probably the fastest candidate for a meteorite dropping event ever observed. This solid meteoroid belonging to the meteor shower survived a significant dynamic pressure of almost 2 MPa and thus ranks among the interplanetary bodies of asteroidal origin that caused the observed meteorite fall. Although a similar Geminid event has been previously presented in the literature, we demonstrate here that this claim was flawed.} 


\authorrunning{P. Spurn\'y et al.}
\titlerunning{Meteorite dropping Geminid}
\maketitle
%

\section{Introduction}

\label{sint}

\subsection{The Geminids and their significance}

The Geminid meteor shower is one of the most active of the year with a stable peak activity regularly observed around 
December 14 \citep{jen06, pec99}. Apart from numerous small particles which cause radar and visual meteors, Geminids are also rich in larger meteoroids which produce bright fireballs and according to their sizes and structure they can penetrate relatively deep in the atmosphere. For decades meteor astronomers have accepted the evident fact that Geminids are more cohesive than any other shower meteors \citep{hal88,spu93}. They ablate rather like solid asteroidal meteoroids than as fluffy cometary dust balls. Also their light curves have very often regular symmetric shapes and are mostly without noticeable flares and sudden terminal decrease. In 1983 the parent body of Geminids was discovered \citep{whip83}. It is the asteroid-comet transition object 3200 Phaethon (its diameter is 5.1 km as found by \citet{han16}), the nature of which is subject to continuous debate. While some authors declare that it is a small asteroid with origin in the main belt \citep{lic07}, a group of at least similar size claims that it is a cometary body. As it is shown in \citet{leo10}, spectral and dynamical properties of Phaethon support its asteroidal origin and connection to one of the largest main belt asteroids, Pallas. On the other hand, from modeling of the formation of Geminid meteor stream seems unlikely to be of asteroidal origin, whereas cometary models described in \cite{rya07}, and also in \cite{rya16, rya18} are in very good agreement with the observed structure of the stream. Similarly \citet{jew10} interpreted sudden short increase of the Phaethon brightness near perihelion observed by NASA's STEREO spacecraft that Phaethon is a "rock comet" in which dust is produced by the thermal fracture and decomposition cracking of hydrated minerals at the high surface temperatures experienced when close to the Sun. 
From present knowledge it is evidently difficult to solve this crucial question about the Geminid-Phaethon origin. 
The best way which would very probably solve this problem is a sample return mission which would bring material from Phaethon 
to the lab. It already happened for comet 81P/Wild 2 \citep[Stardust mission,][]{Sandford2021} and for asteroid 25143 Itokawa \citep[Hayabusa mission,][]{Yoshikawa2015}. Phaethon would be certainly also very suitable target for such mission, but these missions are extremely expensive and still on the edge of our technical skills and possibilities. 
However, recently Phaethon has been selected as the primary target for the upcoming JAXA mission DESTINY$^+$ \citep{Ozaki2022}.
This mission is designed to conduct a close flyby, facilitating high-resolution imaging of the surface and in-situ analysis of the surrounding dust environment. In light of this, the detailed characterization of the physical properties of Geminid meteoroids via ground-based observations acquires renewed urgency. Constraining the bulk density, structure, and ablation behavior of the Geminids is critical, as these parameters provide essential boundary conditions for modeling the dust ejection mechanisms and interpreting the surface evolution data to be returned by the spacecraft \citep{masiero2019}. The synergy between the remote sensing of the meteoroids and the direct observation of the parent body is thus fundamental to resolving the paradoxical nature of Phaethon, which exhibits both asteroidal and cometary characteristics. 

However, if we accept a direct genetic link between the parent body Phaethon and the Geminid meteoroids, and the fact that
the DESTINY$^+$ mission is not designed to collect and transport samples from Phaethon to Earth, we should also test the scenario that these samples could reach Earth on their own. That is, whether there is a possibility that a Geminid meteoroid could survive the flight through the atmosphere and fall to Earth as a meteorite.

\subsection{Upper limits of meteoroid entry velocity for meteorite survival}

The ability of a meteoroid to survive atmospheric entry and produce meteorites is governed by the combined effects of aerodynamic loading, thermal ablation, fragmentation processes, and intrinsic material strength. Although entry velocity plays a fundamental role in controlling these processes, it does not constitute an absolute limiting parameter by itself. Instead, survival is primarily determined by whether individual fragments are able to withstand the dynamic pressure experienced during atmospheric flight and decelerate to subsonic velocities before complete ablation occurs.

Classical meteor physics established that fragmentation is triggered when the aerodynamic (dynamic) pressure exceeds the mechanical strength of the meteoroid \citep{Ceplecha1976}. Subsequent work demonstrated that fragmentation is often a progressive and hierarchical process rather than a single catastrophic disruption, with fragmentation behavior strongly dependent on internal structure and strength heterogeneity \citep{Ceplecha1993}. Because dynamic pressure scales with $\rho v^2$, higher entry velocities generally lead to earlier fragmentation at higher altitudes and increased mass loss. However, this framework implies a probabilistic, rather than absolute, upper limit for meteorite survival.

Instrumentally observed meteorite falls support this interpretation. Most recovered meteorites originate from meteoroids with relatively low to moderate pre-atmospheric velocities, typically below 20~km~s$^{-1}$. For example, the Winchcombe carbonaceous chondrite entered the atmosphere at $\sim$13.9~km~s$^{-1}$ and survived due to low peak dynamic pressures acting on a mechanically weak body \citep{McMullan2023}. Similarly, the \v{Z}\v{d}\'ar nad S\'azavou meteorite fall demonstrates that meteoroids entering at intermediate velocities of $\sim$22~km~s$^{-1}$ can still produce meteorites when fragmentation proceeds gradually and a sufficient fraction of the mass decelerates at low altitudes \citep{Spurny2019}.

At higher velocities, meteorite survival becomes increasingly rare but is not excluded. The Maribo CM2 meteorite fall represents one of the most extreme documented cases, with a pre-atmospheric velocity of $28.3 \pm 0.3$~km~s$^{-1}$ \citep{bor19}. Detailed modeling showed that most of the mass was disrupted at high altitudes; nevertheless, a small fraction of mechanically stronger fragments survived dynamic pressures of approximately 3--5~MPa, decelerated efficiently, and reached terminal velocity at sufficiently low altitudes. This case highlights the crucial role of material heterogeneity and differential fragment strength in enabling survival at unusually high entry velocities.

An even higher entry velocity was documented for the Sutter's Mill meteorite fall, with an initial speed of approximately $28.6 \pm 0.7$~km~s$^{-1}$ \citep{Jenniskens2012}. Despite this extreme velocity and extensive fragmentation, small meteorite fragments were recovered. Detailed analyses revealed that survival was facilitated by early fragmentation into a large number of small bodies, some of which retained sufficient strength to decelerate rapidly and avoid complete ablation. Together with Maribo, Sutter's Mill establishes the current empirical upper bound for confirmed meteorite-producing fireballs at velocities close to 29~km~s$^{-1}$.

In this context, meteoroids associated with the Geminid meteor shower represent an even more extreme regime. Geminids enter the atmosphere at velocities of approximately 35--36~km~s$^{-1}$, significantly exceeding those of all known meteorite-producing fireballs. At such velocities, dynamic pressures and ablation rates are expected to be extremely high, leading to rapid and efficient destruction of even mechanically strong material. Consequently, the survival of Geminid meteoroids to produce meteorites is generally considered highly improbable. Nevertheless, by analogy with the Maribo and Sutter's Mill events, the possibility cannot be ruled out entirely that exceptionally rare, compact, and mechanically strong fragments could survive a favorable combination of entry angle, fragmentation history, and deceleration.

The recovery of even a very small fragment of Geminid material would have profound scientific implications. As was discussed above, Geminids are dynamically associated with asteroid (3200)~Phaethon, whose nature and relationship to meteoroids remain subjects of active debate. Laboratory analysis of a Geminid-derived meteorite would provide an unprecedented opportunity to directly constrain the physical, chemical, and isotopic properties of this unique meteoroid population and to establish a direct link between meteoroids and their parent body. Although such an event would be extraordinarily rare, its potential scientific value justifies continued interest in deep-penetrating Geminid fireballs.

\begin{figure*}
\centering
\includegraphics[width=\textwidth, height=11.25cm, keepaspectratio]{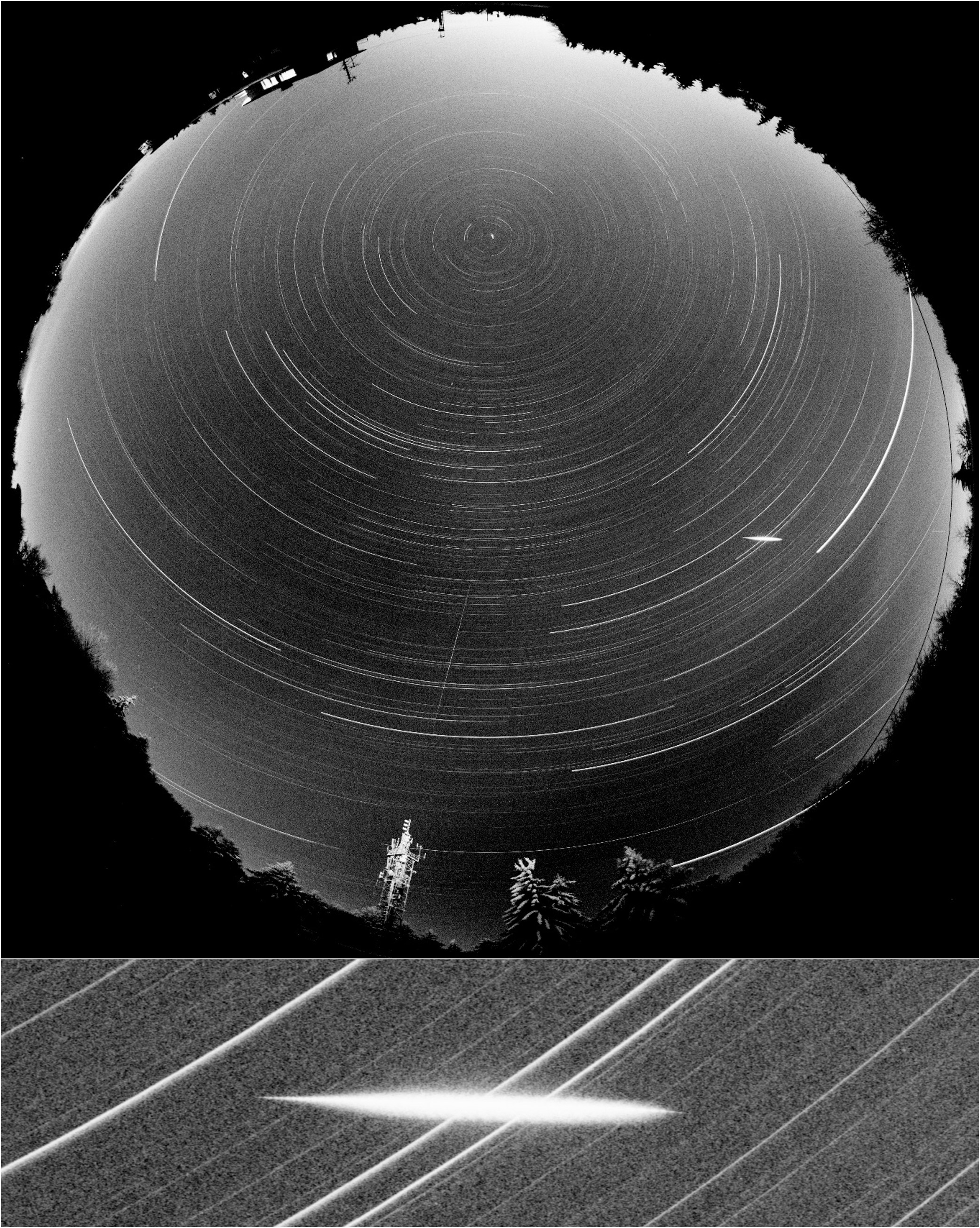}  
\hfill 
\includegraphics[width=\textwidth, height=11.25cm, keepaspectratio]{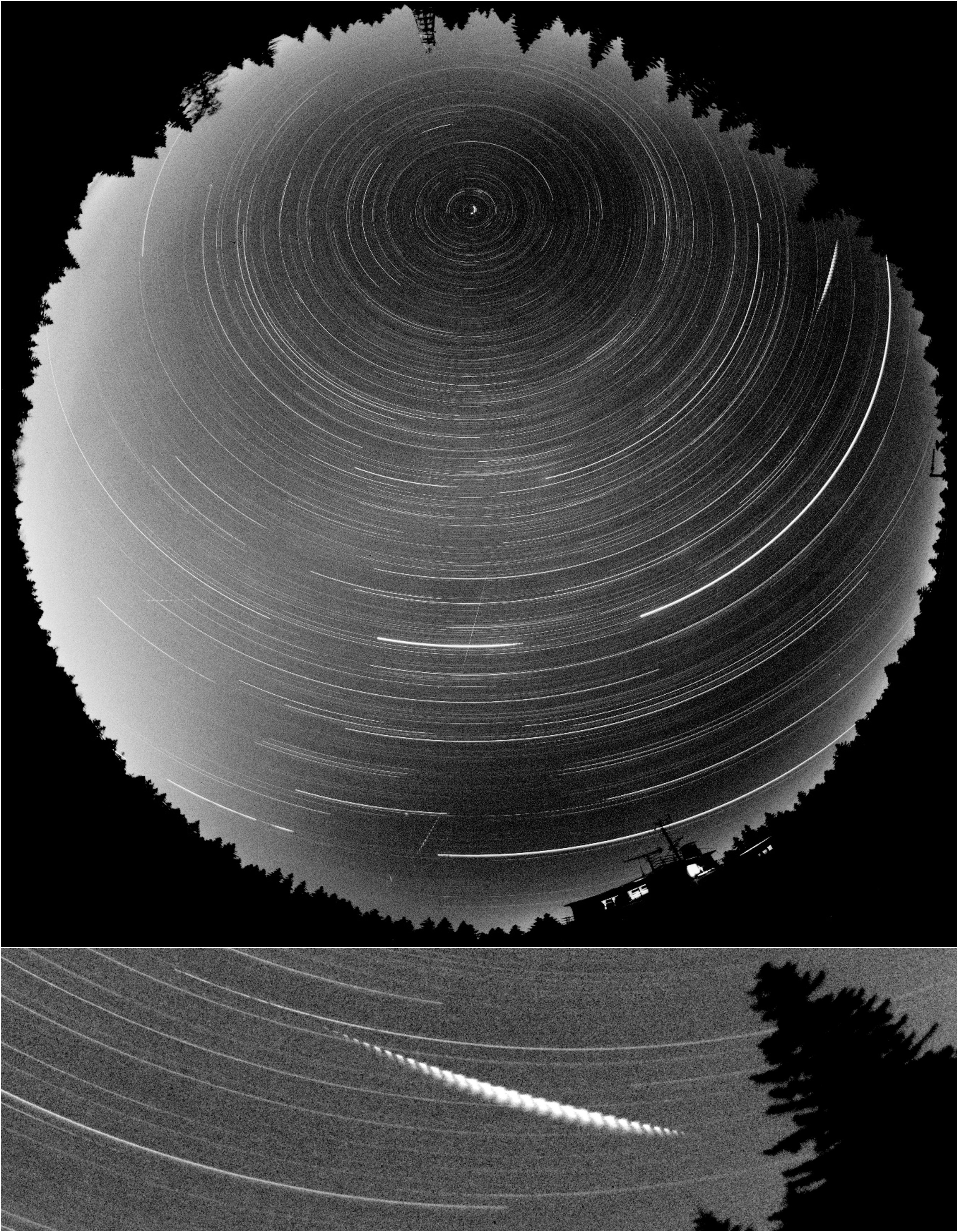}  
\caption{All-sky and detailed images of the EN131212\_041258 Geminid fireball from both stations. (\textit{a}) Left - station  P\v{r}imda, (\textit{b}) Right - station Chur\'{a}\v{n}ov.}
\label{bothimages}
\end{figure*}

\begin{figure}
    \centering
    \includegraphics[width=1\linewidth]{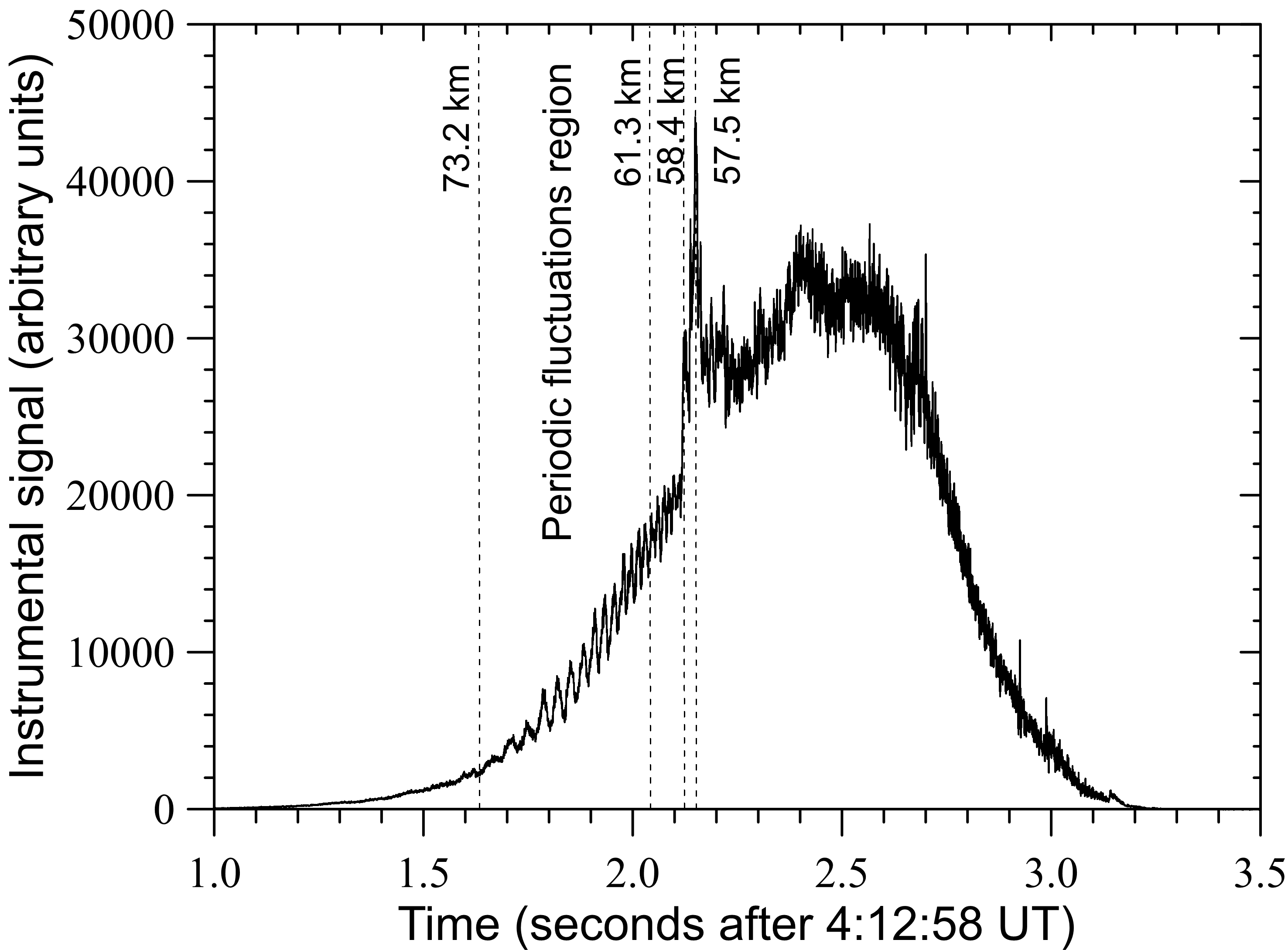}
    \caption{Radiometric curve obtained by AFO at the  P\v{r}imda station. The marked heights (dashed lines) delimit the area of regular brightness fluctuations (see section 3.3) and also heights of two consecutive flares, which correspond to fragmentation events described in section 3.4.}
    \label{LCint_11fin}
\end{figure}

From all of the about mentioned aspects every new reliable instrumental data either about Geminids or Phaethon itself can help 
to reveal at least partially the mystery of the origin of the Geminids and their parent body Phaethon. This is also the main purpose of this paper, where we clearly give an answer to one important question, which was only a matter of speculation up to now - can Geminids produce meteorites? We present here observational evidence of one extraordinary deep penetrating Geminid meteoroid, which survived severe deceleration and penetrated much more deeply than any previous instrumentally observed Geminid with surprisingly low terminal speed. It implies that under some favorable conditions and circumstances it is even for such 
fast meteoroids possible to survive the atmospheric flight and land on the ground. And it could happen that, thanks to a combination of other favorable circumstances, the landed fragment of the Geminid meteor shower will be found. Although it did not happen this time, it is not excluded in future that with more favorable circumstances such as especially larger initial mass with similar structure and atmospheric passage conditions like for the presented case, some Geminid meteorite, i.e. small piece of Phaethon, can be recovered and studied in a laboratory. It is evident that such Geminid sample would have an invaluable scientific value. 

\begin{table*}[t]
\caption{Locations of the cameras, their distances to the fireball beginning and end, span of the recorded heights, elevations, and total recorded length.}
\label{camerastable}
\centering
\begin{tabular*}{\textwidth}{@{\extracolsep{\fill}} lll|llc|cc|cc|cc|c}
\hline
\noalign{\smallskip}
\multirow{2}{*}{Station} &{Network}& \multirow{2}{*}{Camera} & \multicolumn{3}{c|}{Coordinates (WGS84)} & \multicolumn{2}{c|}{Distance (km)} & \multicolumn{2}{c|}{Height (km)} & \multicolumn{2}{c|}{Elev. ($^\circ$)} & Length \\
 &No.&& Longit. $^\circ$E	& Latitude $^\circ$N & h (m) & Beg	& End	& Beg	& End  & Beg & End & (km)\\
\noalign{\smallskip}
\hline
\noalign{\smallskip}
 P\v{r}imda & EN 11 & AFO &	12.67790 & 49.66936	& 745 & 128.04 & 46.76 & 99.33 & 32.54 & 50.0 & 42.7 & 81.90	\\
Chur\'{a}\v{n}ov & EN 4 & AFO & 13.61495 & 49.06843 & 1119 & 177.35 & 125.22 & 90.52 & 34.74 & 28.7 & 15.2 & 68.42  \\
\noalign{\smallskip}
\hline
\noalign{\smallskip}
\end{tabular*}
 \tablefoot{The abbreviations used - EN means European Fireball Network, AFO means Autonomous Fireball Observatory, which produces all-sky photographic images and radiometric light curves. The presented solution assumes straight fireball trajectory.}
\end{table*}

\section{Instruments and Observational data} 
As mentioned above, this paper presents data on an extraordinary fireball clearly belonging to the Geminid meteor shower, which was recorded by instruments in the Czech part of the European Fireball Network. This fireball, which flew on December 13, 2012, at 4:12:58.7 UT (time is valid for the photographic beginning of its luminous atmospheric trajectory), was very well recorded photographically and radiometrically at two stations, namely  P\v{r}imda and Chur\'{a}\v{n}ov which are located in the western part of the network. Fortunately, these were the closest EN stations to the fireball's trajectory. Their exact location, the distances between the beginning and end of the recorded trajectory, the range of altitudes, and the total recorded length are listed in  Table \ref{camerastable}. At that time, these stations were equipped with analog autonomous fireball observatories (AFO) that took all-sky photographs of fireballs as shown in Figure~\ref{bothimages} and, at the same time, recorded the detailed light curves of fireballs with radiometers which are part of each AFO with a very high time resolution of 5000 samples per second as shown in Figure~\ref{LCint_11fin}. This system is described in detail in \citet{spu07}.   
Both stations had a perfectly clear sky, not only at the time of the fireball passage, but for almost all of this night, 
including the beginning and end of the exposure. This was important for two main reasons. First, the transparent sky enables 
us to measure more details on the fireball luminous path including breaks (time marks on the luminous path enabling velocity 
determination). Perfect sky conditions also significantly helped in the correct reduction of both images because we had enough reference stars to precisely define the coordinate system on each image and determine the coordinates of each measured point on the fireball luminous trajectory. This was a very important factor because in case of only two stations the correct reduction is crucial point (the reduction procedure of these photographic images is described in \cite{bor95}). Fortunately, the geometry was relatively good, the fireball was well above the horizon on both stations, especially on  P\v{r}imda where angular heights above the horizon varied in the range from  50.0$\degr$ to 42.7$\degr$. The situation on the Chur\'{a}\v{n}ov image was a bit worse but the fireball was still 28.7$\degr$ above the ideal horizon for the beginning, and 15.2$\degr$ for the end. This case
completely proved our methodology in using large format fish-eye lenses (Zeiss Distagon 3.5~/~30~mm) in combination with photographic film (Ilford FP4 sheet film 9 $\times$ 12 cm, one per night). Thanks to the precise imaging and high resolution of both images (we scan and measure each point with precision of 5~$\mu$m) we were able not only to reduce both images but, as we outline below, also to determine the atmospheric trajectory and dynamics with very high degree of accuracy.

\section{Results}

\subsection{Atmospheric trajectory}

\begin{figure}
\centering
\includegraphics[width=\hsize]{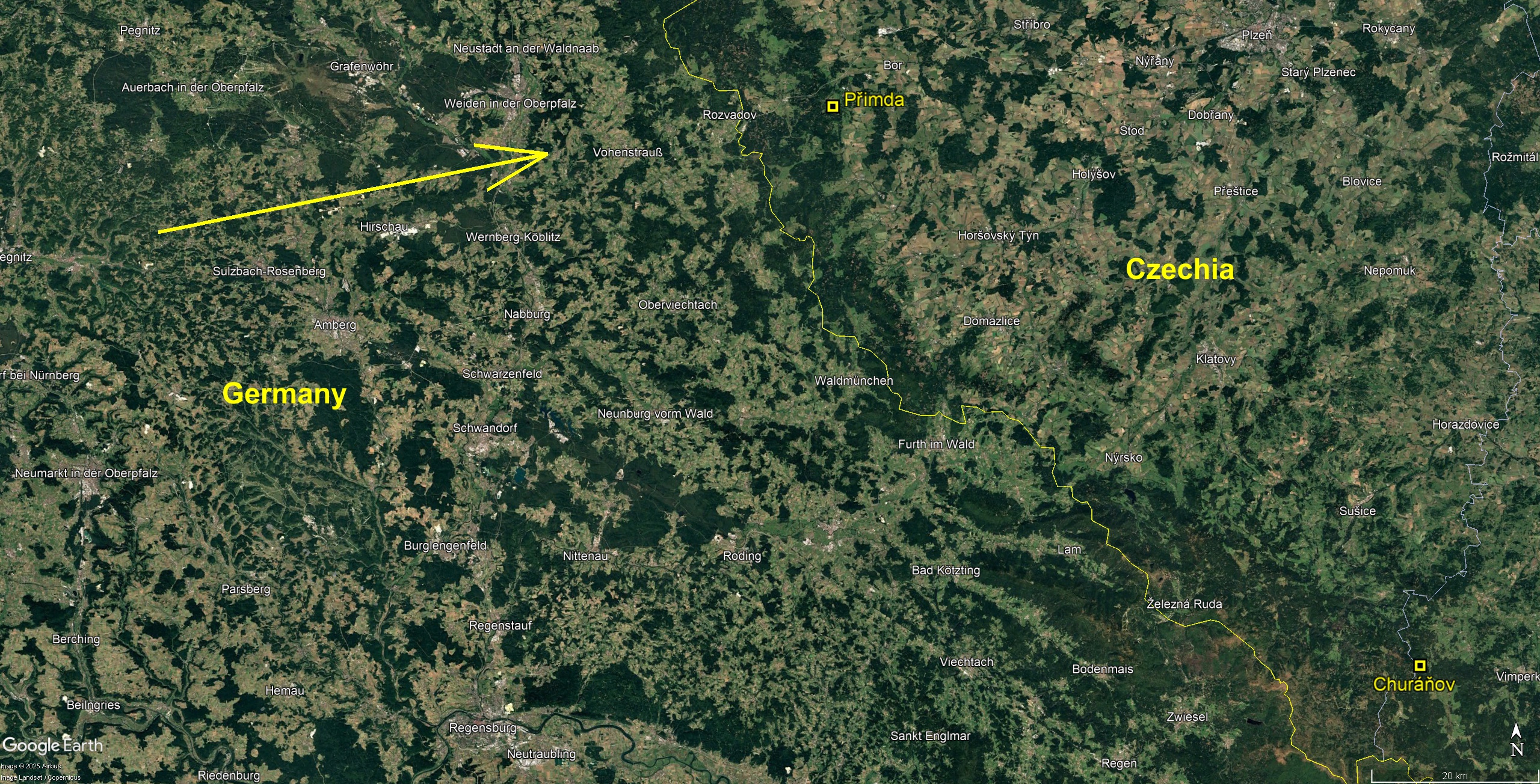}
\caption{Ground projection of the recorded atmospheric trajectory (yellow arrow) of the EN131212\_041258 Geminid fireball and location of the Czech stations  P\v{r}imda and Chur\'{a}\v{n}ov where its records were taken. The average slope of the atmospheric trajectory was 54.7\degr  and its length was 81.9 km. (source of the map: Google Earth)}
\label{atmtraj+st}
\end{figure}

The bolide trajectory was computed by the least-squares method of \cite{bor90}. The trajectory was first assumed to be
straight. Corrections for curvature due to gravity were applied at the end, when the linear trajectory and velocity were known.
Since the fireball was observed from only two stations, it is evident that trajectory solutions will almost always yield a mathematical fit, though such solutions may not necessarily reflect reality. Reliability is increased by the geometry of the stations' positions relative to the fireball, i.e., the angle of the fireball's planes from each station at which they intersect, which in this case was $49.5^{\circ}$. Given the high quality and resolution of both images, this value of the intersection angle is entirely sufficient. Furthermore, the distance of the fireball from both stations plays also an important role (provided for the initial and terminal points along with other parameters in Table~\ref{camerastable}). Similarly, for reliable reduction, its elevation above the horizon and a sufficient number of reference stars are particularly important for this type of all-night and all-sky fixed images. Ideally, reference stars should be distributed not only across the entire sky, but also around the fireball, to prevent uncertainties arising from the extrapolation of positions at larger zenith distances. All these criteria were satisfied for this fireball, indicating that the determination of the atmospheric trajectory should be reliable even it is only from two stations. In addition, another very important check is the determination of the speed along the trajectory independently determined from both stations. And again, in this case, the speeds derived from both stations matched very well. Although the breaks could not be measured in the middle part of the fireball in the P\v{r}imda image because of low angular speed and high brightness, the initial and terminal phases could be measured well. From all of the above, it can therefore be concluded that the determined atmospheric trajectory, as shown in the Table~\ref{atmtrajtable} and illustrated in the Figure~\ref{atmtraj+st} is highly reliable, both in terms of spatial location and dynamics. This can also be seen from the standard deviations of the individual values listed in this Table~\ref{atmtrajtable}.

\begin{table}[h]
\caption{Atmospheric trajectory data of the EN131212\_041259 Geminid fireball.}
\label{atmtrajtable}
\centering
\begin{tabular*}{\columnwidth}{l| @{\extracolsep{\fill}} |r @{ $\pm$ } l r @{ $\pm$ } l}
\hline
\noalign{\smallskip}
Parameter & \multicolumn{2}{c}{Beginning} & \multicolumn{2}{c}{Terminal} \\
\noalign{\smallskip}
\hline
\noalign{\smallskip}
Height (km)             & 99.313    & 0.006      & 32.531    & 0.004   \\
Longitude ($^\circ$ E)  & 11.57424 & 0.00007   & 12.21040 & 0.00005  \\
Latitude ($^\circ$ N)   & 49.53841 & 0.00006   & 49.62084 & 0.00004  \\
Slope ($^\circ$)        & 54.832   & 0.007     & 54.549   & 0.007    \\
Azimuth ($^\circ$)      & 78.488   & 0.012     & 78.972   & 0.012   \\
Speed (km s$^{-1}$)  & 35.75    & 0.06      & 6.8      & 0.3       \\
L (km) / T (s)      & \multicolumn{4}{c}{81.89 / 2.79}       \\
\noalign{\smallskip}
\hline
\end{tabular*}

\tablefoot{L is the length and T is the duration of the recorded trajectory; azimuth is measured from the south. The heights and slopes include the expected effect of the trajectory bending by gravity.}

\end{table}

The beginning and end points given in Table~\ref{atmtrajtable} are the points at which the bolide began and ceased to be visible in the records. Here it is important to note that the sensitivity of the analog AFOs, which we used at the time and which took all-sky images on large-format (9$\times4$12 cm) flat films, was approximately $-3$ to $-4$ apparent magnitude. It is significantly lower sensitivity than that of current digital recordings, as described, for example, in \cite{Spu17} or \cite{bor22}. This means that with more sensitive cameras or video cameras as we use now (since 2014), both the beginning and the end of the bolide would most likely have been recorded significantly earlier or later, respectively. This effect is usually more pronounced for the beginning than for the end of a bolide, which was certainly the case with this bolide. We can see it, for example, in Figure~\ref{LCint_11fin} or Figure~\ref{lc-model-h}, where it is clear that the brightness at the beginning of the bolide increased much more gradually, in contrast to the steeper decrease at the end. 
The exact time corresponding to the observed beginning of the bolide shown in the Table~\ref{atmtrajtable} is $4^{\rm h}12^{\rm m}58\fs74$~ UT. The ground projection of the atmospheric trajectory is shown in Figure~\ref{atmtraj+st}. This trajectory was slightly steeper than for an average fireball with an average slope of 54.7 degrees, and its luminous path length was 81.90 km. The entire bolide flew over the eastern part of Germany (Bavaria) near the border with the Czech Republic. As can be seen from the fragmentation model (see Section 4), only a small remnant of the original meteoroid, weighing a few grams at most, could have survived the ablation phase of flight and fallen to the Earth's surface. According to the calculation of the dark flight trajectory, the theoretical impact site is also still in Germany, near the village of Moosbach, only about 20 km away from the  P\v{r}imda station, where the bolide was recorded by our instruments. However, such a small estimated size of the meteorite essentially rules out a systematic search in the conditions of central Europe.   

\begin{table}
\caption{Radiant and heliocentric orbit of the Geminid fireball EN131212\_041259.} 
\label{orbit} 
\centering
\begin{tabular}{l c c c} 
        \hline 
        Parameter & Unit & Fireball & Geminids \\ 
        \hline
        $\alpha_{\mathrm{R}}$ & (deg) & $114.457 \pm 0.011$ \\ 
        $\delta_{\mathrm{R}}$ & (deg) & $33.220 \pm 0.005$ \\ 
        $v_{\infty}$ & $\mathrm{(km\,s^{-1})}$ & $35.75 \pm 0.06$ \\ 
        $\alpha_{\mathrm{G}}$ & (deg) & $113.188 \pm 0.011$ & 113.0 \\ 
        $\delta_{\mathrm{G}}$ & (deg) & $32.433 \pm 0.006$ & 32.4 \\ 
        $v_{\mathrm{G}}$ & $\mathrm{(km\,s^{-1})}$ & $34.16 \pm 0.07$ & 33.8 \\ 
        $v_{\mathrm{H}}$ &$\mathrm{(km\,s^{-1})}$ & $33.58 \pm 0.04$ \\ 
        $a$ & (A.U.) & $1.315 \pm 0.005$ & 1.31 \\ 
        $e$ & -- & $0.8930 \pm 0.0008$ & 0.889 \\ 
        $q$ & (A.U.) & $0.1407 \pm 0.0005$ & 0.145 \\ 
        $Q$ & (A.U.) & $2.490 \pm 0.011$ \\ 
        $\omega$ & (deg) & $324.77 \pm 0.03$ & 324.3 \\ 
        $\Omega$ & (deg) & $261.3777$ & 261.6 \\ 
        $i$ & (deg) & $23.67 \pm 0.09$ & 22.9 \\ 
        $P$ & (years) & $1.508 \pm 0.009$ \\ 
        $T_{\mathrm{J}}$ & -- & $4.370 \pm 0.016 $ \\ 
        \hline
\end{tabular}
\tablefoot{
        Quantities with indices R, G, and H refer to the observed radiant, geocentric, and heliocentric velocities, respectively.The parameters $a, e, q, Q, \omega, \Omega, i$ represent the semi-major axis, eccentricity, perihelion distance, aphelion distance, argument of perihelion, longitude of the ascending node, and inclination, respectively. $v_{\infty}$ is the velocity before atmospheric entry. $P$ is the orbital period. $T_{\mathrm{J}}$ is the Tisserand parameter with respect to Jupiter, which is used to classify the orbit type of the object. The last column are the median Geminid values from \citet{JennBook2}. All angular entries are given for standard equinox J2000.0.
    }
\end{table}

\subsection{Heliocentric orbit}

The geocentric radiant ($\alpha _G, \delta_G$) and heliocentric orbit were computed from the apparent radiant ($\alpha _R, \delta_R$) and entry velocity  ($v_\infty$) by the analytical method of \cite{cep87}, with a small modification accounting for the trajectory curvature described in \cite{bor22}. The entry speed was determined from the dynamic fit of all breaks from both records (altogether 26) along the first 43 km of the fireball length (at heights above 64 km), where the fireball was measured well and severe fragmentation had not yet begun. The initial mass was set to the value obtained from light-curve modeling (see Section 4).
The apparent and geocentric radiant and orbital elements (all for standard equinox J2000.0) are given in Table~\ref{orbit}. The main and very important result of the heliocentric orbit analysis is that, according to all these orbital parameters from Table~\ref{orbit}, it is indisputable that the observed bolide belongs to the Geminid meteor shower. 

\subsection{Radiometric light curve and flickering}

As mentioned above, the detailed light curve was recorded by the brightness sensors (radiometers) at all AFOs which were in operation that night in our network. The best records, i.e. the records where the signal to noise ratio was the highest, were taken from the stations  P\v{r}imda and Chur\'{a}\v{n}ov. This observed light curve from AFO at  P\v{r}imda is shown in Figure~\ref{LCint_11fin}. From the photographic images we know that there was a clear sky during the fireball passage at both stations so these radiometric records were not affected or distorted by clouds. The differences in the recorded radiometric curves are only due to the different distances from the bolide and, therefore, to the different intensities of the recorded light signals. In reality (i.e., in absolute values), both records are practically identical, which can be clearly seen for example in Figure~\ref{lc-model-h}, where the light curves are given in absolute magnitudes, i.e., normalized to a unit distance of 100 km.

\begin{figure}
    \centering
    \includegraphics[width=1\linewidth]{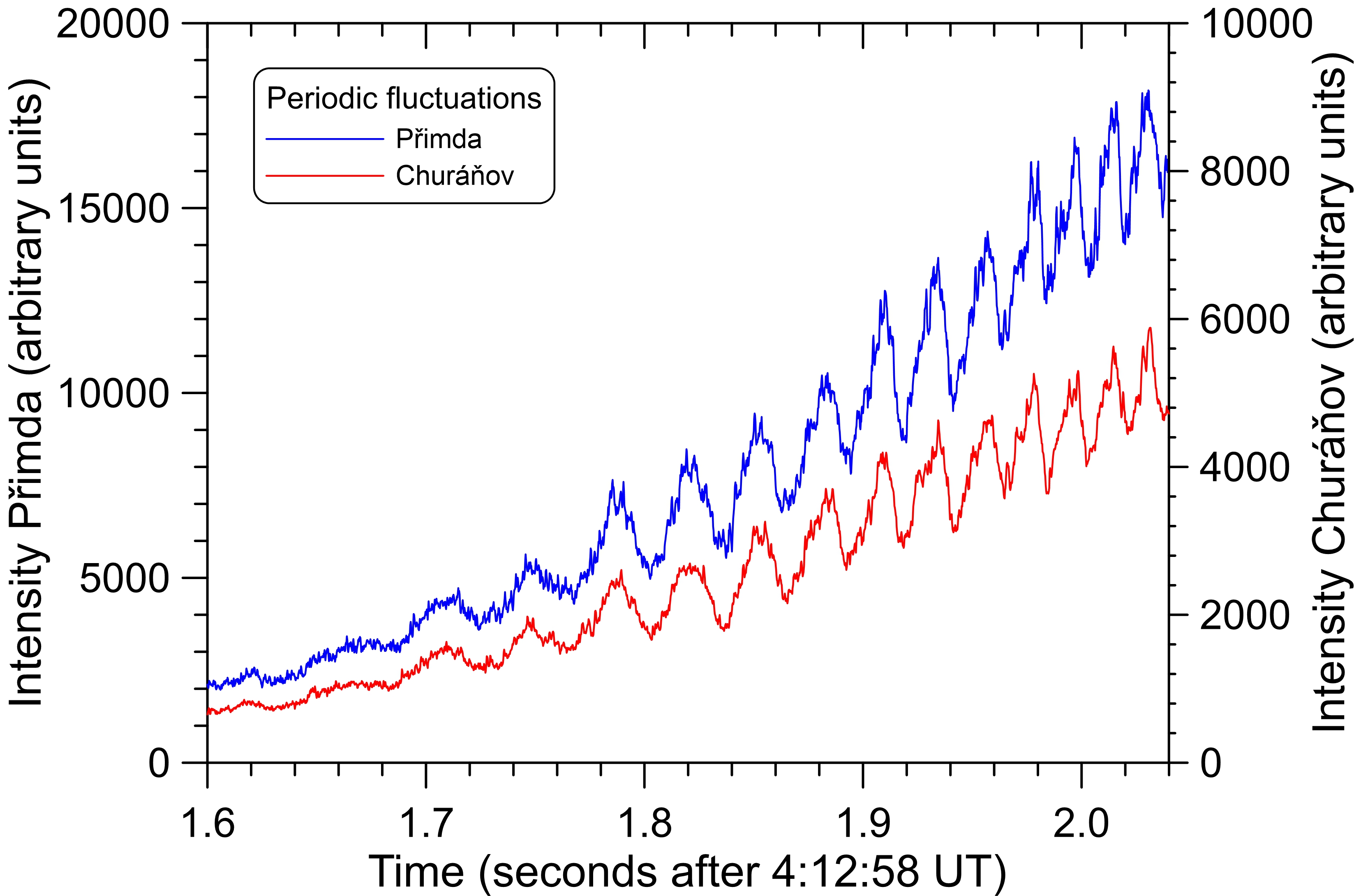}
    \caption{Detailed section of the radiometric light curve of the EN131212\_041259 Geminid fireball recorded independently at both stations, showing the area of periodic fluctuations corresponding to altitudes between 73.2 and 61.3 km, as indicated in Figure~\ref{LCint_11fin}}
    \label{LCflicker}
\end{figure}

\begin{figure}
    \centering
    \includegraphics[width=1\linewidth]{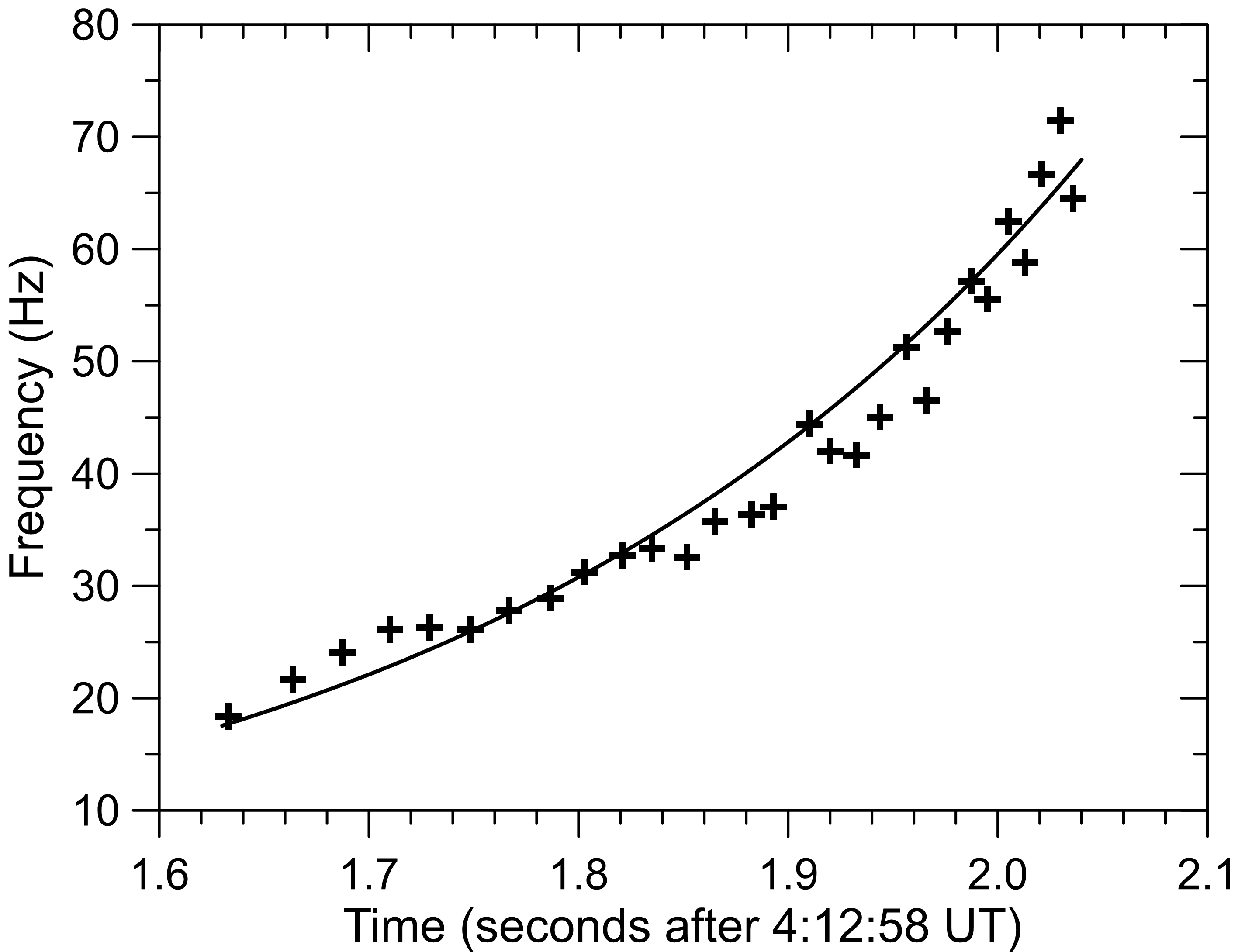}
    \caption{Frequency of periodic fluctuations as a function of fireball time. The range of values corresponds to heights from 73.2 km to 61.3 km. It is evident that the frequency increases with time, and this trend is approximately exponential.}
    \label{Flickfreqfin}
\end{figure}

Overall, it can be said that this is a relatively typical light curve for the Geminids, i.e., relatively symmetrical with one significant brightening. Upon closer examination, however, the behavior in the initial phase before the main flare is particularly interesting. Here, during the gradual brightening, clearly visible regular fluctuations in brightness with increasing frequency appear. This is clearly shown in Figure~\ref{LCflicker}, where this part of the light curve is enlarged for both stations, i.e.,  P\v{r}imda and Chur\'{a}\v{n}ov. This comparison clearly shows how well the two independent records taken from locations 95 km apart correspond to each other. This is therefore certainly a real representation of the bolide's brightness. If we plot the minimum and maximum times of these fluctuations, as it can be seen in Figure~\ref{Flickfreqfin}, their frequency increases approximately exponentially from the original value of just under 20 Hz to a value of around 70 Hz. This is at an altitude of 61.3 km, when these fluctuations cease to be clearly distinguishable. It is shortly before the main brightening, which begins at an altitude of 58.4 km, as seen in Figure~\ref{LCint_11fin}. In the area between these altitudes, further brightening becomes more pronounced and it is no longer clear what belongs to the original periodicity.  

In terms of the details of the light curve, the part after the maximum is also interesting. The large concentration of small but very rapid and irregular changes in brightness practically until the end of the fireball's luminosity is not caused by noise, but by real millisecond pulsations (in the order of a few milliseconds at most), which resemble sparking. Especially for the recording from  P\v{r}imda, the signal-to-noise ratio is very high practically throughout the entire trajectory, except for the very beginning (the first 0.5 seconds) and the very end (the last 0.3 seconds). Similarly, the recording from Chur\'{a}\v{n}ov in the brighter phase (approximately above magnitude -5) perfectly copies all these minor phenomena recorded from  P\v{r}imda. It is therefore beyond doubt that all these minor rapid brightenings are a real manifestation of the passage of this small meteoroid through the atmosphere. Both the periodic fluctuations with increasing frequency before the maximum given by a more significant brightening, which is related to the fragmentation of the original meteoroid as described in the following section, and the way the bolide manifested itself as "sparkling" after the maximum, are presented here as a real observed fact without us knowing its clear explanation. As for the phase before the maximum, it is possibly caused by the rotation of the original irregularly shaped meteoroid, but this is far from certain. Another possibility is a quasi-periodic removal of the molten layer from the surface of the meteoroid. And as for the short, rapid, and irregular brightening after the maximum, this could be related to the triboelectricity as proposed in the work of \cite{spu08}.  In conclusion, we can state that during the entire period of using radiometers in our bolide network, i.e. since 2004, similar behavior has not been observed as standard for the Geminids, but it is not very rare either, especially for those that are bright and, in terms of duration, longer (usually longer than 2 seconds). 

\subsection{Fragmentation modeling}

The observed fireball light curve showed in the previous section and dynamics were used to study meteoroid atmospheric fragmentation. 
The semi-empirical fragmentation model described in detail in \citet{2strengths} was applied.
The model considers regular fragments, eroding fragments, and dust particles. The fragmentation events and 
their outcome were adjusted manually to fit the light curve and deceleration of the fireball.

\begin{figure}
\centering
\includegraphics[width=\hsize]{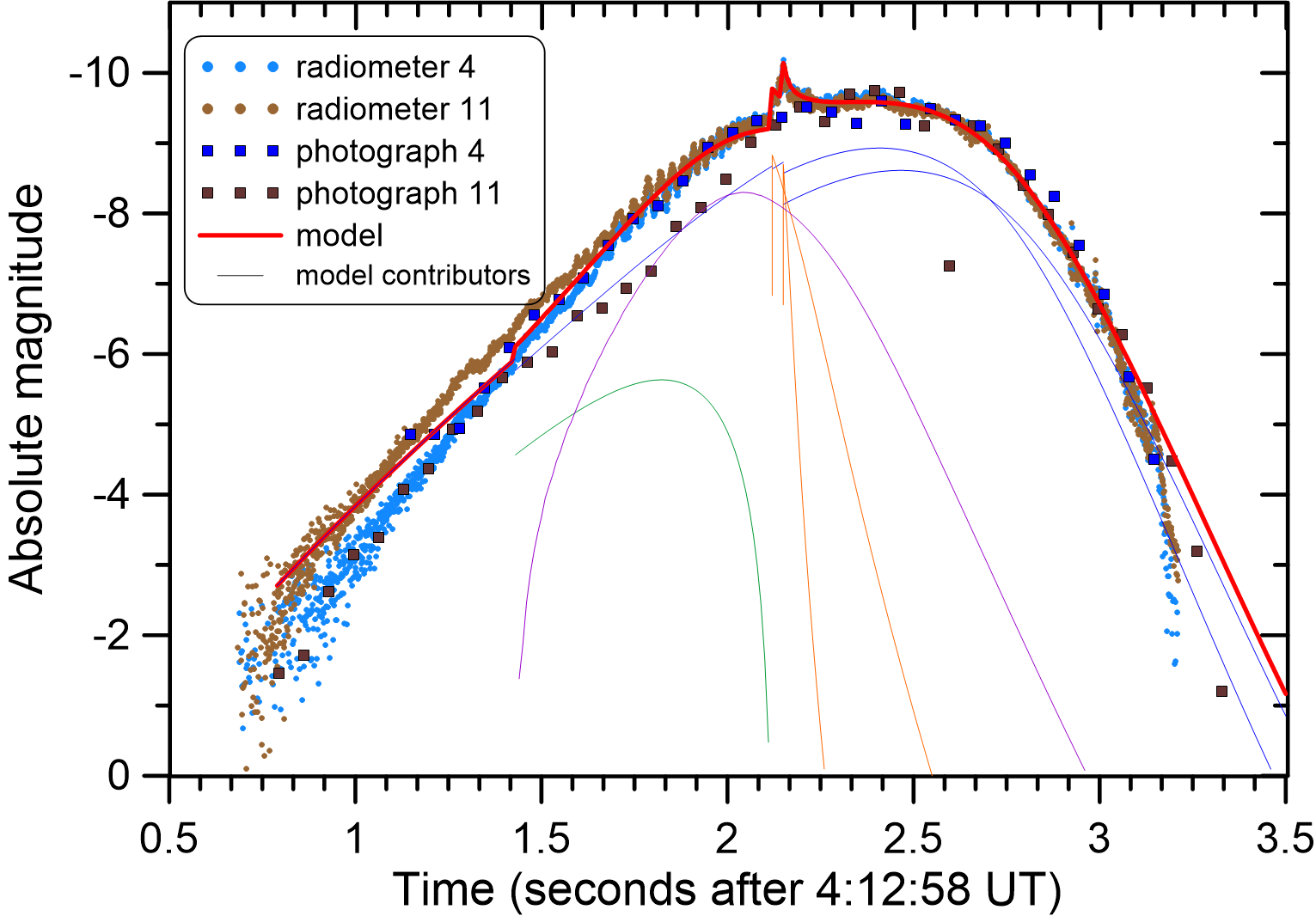}
\caption{Observed and modeled light curve. Individual contributors to the modeled light curve are regular fragments (blue),
eroding fragments (green), eroded dust (purple), and immediately released dust (orange).}
\label{lc-model}
\end{figure}

\begin{figure}
\centering
\includegraphics[width=\hsize]{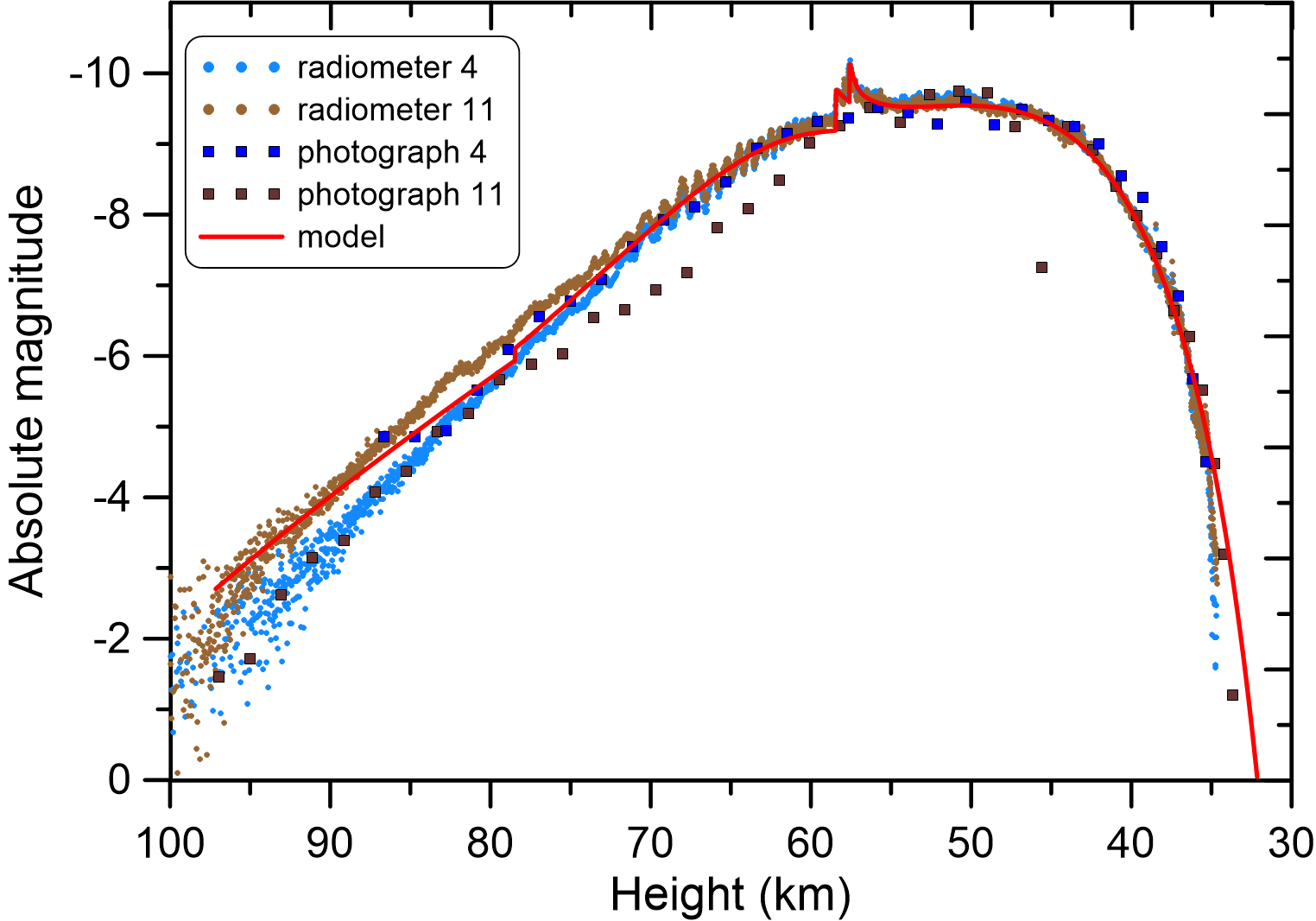}
\caption{Observed and modeled light curve plotted as a function of height.}
\label{lc-model-h}
\end{figure}

Figure~\ref{lc-model} shows the fit of the light curve. Radiometric and photographic photometric data are plotted together
with the model light curve. The contributions of the individual fragments and the dust released during the fragmentation
or from eroding fragments are also shown. Figure~\ref{lc-model-h} shows the same fit but plotted
as a function of height instead of time. For radiometers, heights were assigned to the observing times using the dynamic fit
of the fireball (height as a function of time). Individual contributors are not shown in this plot.

According to the model, the fireball was produced by a meteoroid of mass of 0.25 kg with an initial velocity of 35.75 km s$^{-1}$.
The meteoroid density was set to 2500 kg m$^{-3}$. Using a density higher than 3000 kg m$^{-3}$ resulted in difficulties obtaining a good fit. The other parameters of the meteoroid were $\Gamma A=0.7$ (the product of drag coefficient and shape coefficient)
and $\sigma = 0.005$ s$^2$ km$^{-2}$ (the ablation coefficient). The same parameters were used for daughter 
fragments and dust particles, unless otherwise noted. For the used luminous efficiency, see \citet{2strengths}.

A minor fragmentation was modeled at the height of 78 km. Part of the mass (45 g, i.e.\ 18\%) was separated here
in the form of an eroding fragment. The eroding fragment was needed in order to fit the light curve shape between heights 78--58 km. The masses
of eroded dust particles were $2 \times 10^{-4}$ kg. Note that
regular brightness fluctuations with increasing frequency, well visible in the radiometric curves between heights 
74--58 km, which were discussed in the previous section, were not fitted because they were probably caused by another process than fragmentation.

\begin{figure}
\centering
\includegraphics[width=\hsize]{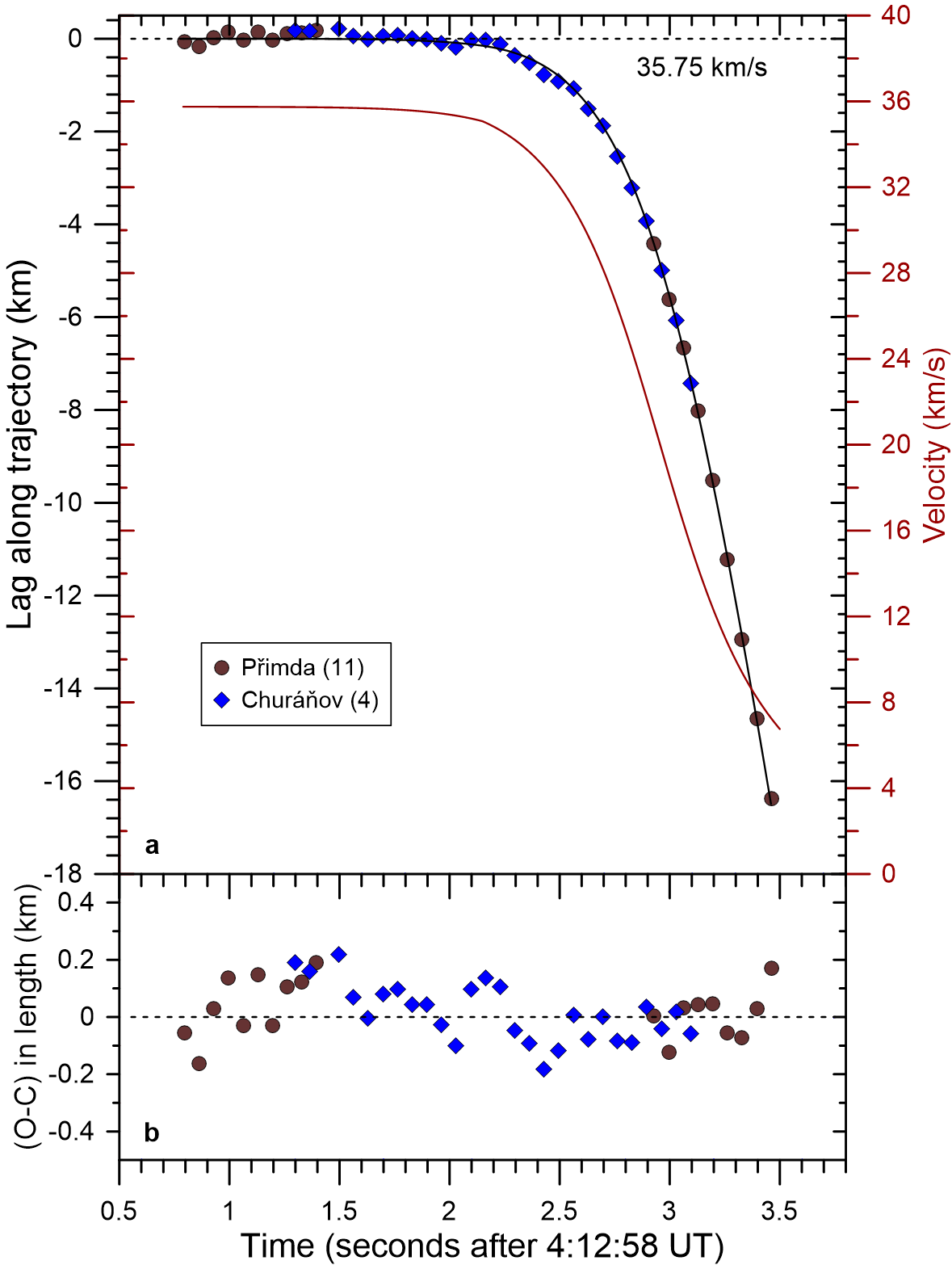}
\caption{Dynamics of the fireball. The upper panel \textbf{a} shows the observed (symbols) and fitted (black line) lag in the
length along the trajectory in respect to a hypothetical body with constant velocity of 35.75~km s$^{-1}$. The red line with
scale on the right shows the velocity derived from the fit. The lower panel \textbf{b} shows the residuals of the observed points
relative to the fit.}
\label{decel}
\end{figure}

\begin{figure}
\centering
\includegraphics[width=\hsize]{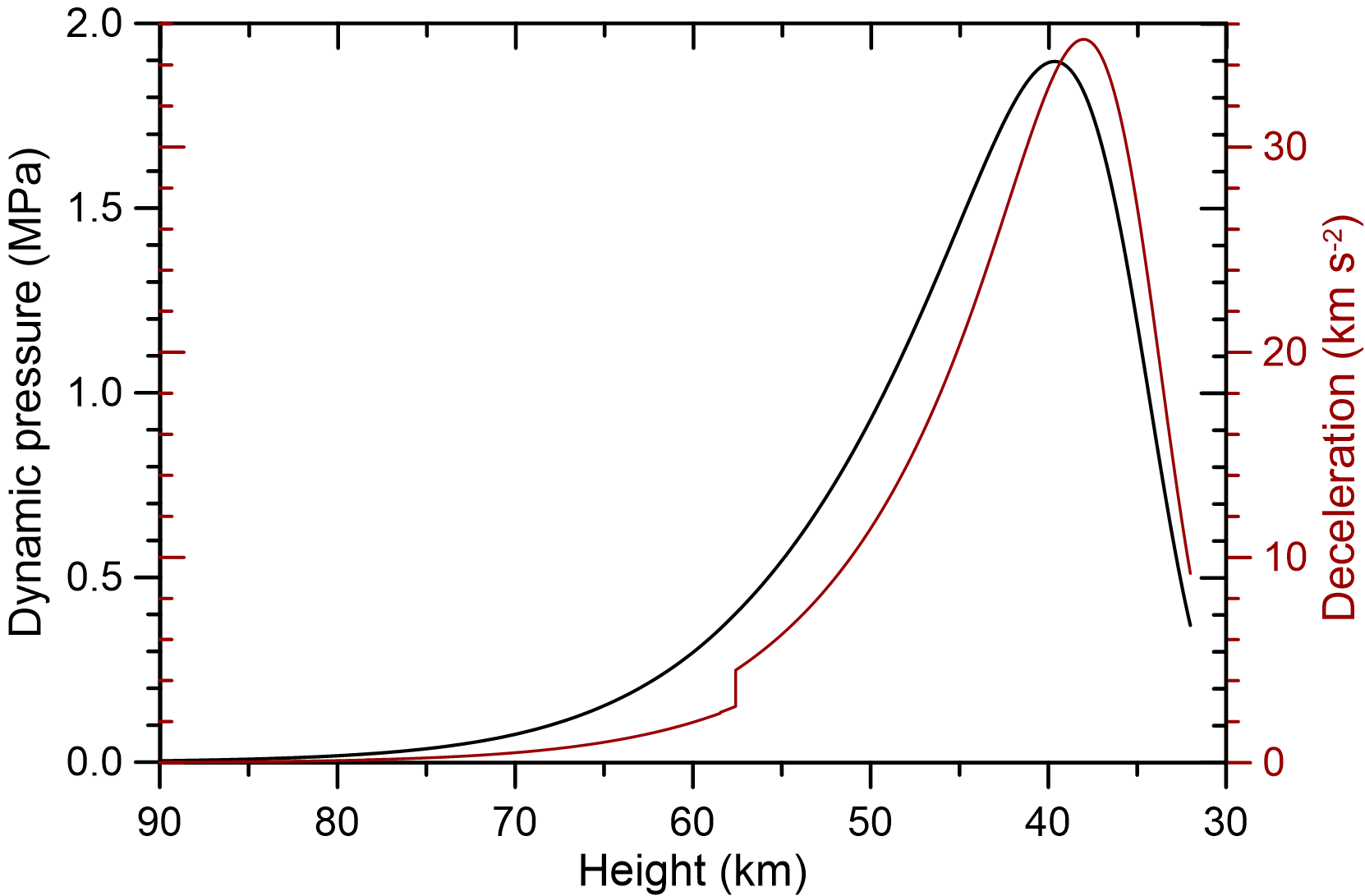}
\caption{Dynamic pressure and deceleration (in red with scale on the right)
as a function of height. The curves were obtained from the dynamic fit presented in Fig.~\ref{decel} and are valid for the largest modeled fragment.}
\label{pressure}
\end{figure}

While the exact height of the first fragmentation is somewhat uncertain, the two consecutive fragmentation events at heights
58.4 km and 57.5 km are well recognizable by flares on the light curve. The flares were produced by released dust particles
of masses of $\sim 10^{-7}$ to $10^{-5}$ kg and a total mass of about 11 g. The meteoroid mass before the fragmentation was 
$\sim$184 g. After the fragmentation, the fireball started to strongly decelerate, as shown in Figs.~\ref{decel} and \ref{pressure}. The mass of the largest fragment could be determined from the dynamics. All dynamic measurement toward the end of the fireball could be fitted by a fragment of initial mass of 39 g. The ablation coefficient had, however, to be decreased to 0.004 s$^2$ km$^{-2}$.
The computed mass at the end of the fireball was 3.7 g. The velocity decreased to 6.8 km s$^{-1}$ at that point.
When extrapolated to the velocity 2.5 km s$^{-1}$, when the ablation usually stops, the mass decreased to 3.4 g. A meteorite of that mass probably landed. Good fit can be obtained also with even lower ablation
coefficient of 0.003 s$^2$ km$^{-2}$, in which case the meteorite mass would be 5 g. We consider this scenario less probable. On the other hand, a mass lower than 3 g at the time of the last velocity measurement is not compatible with dynamic data.

The light curve fit below the height of 57 km required more than one fragment. Another 39 g fragment with the same parameters
and three 31 g fragments with the normal ablation coefficient of 0.005 s$^2$ km$^{-2}$ were added to reproduce the general shape of the light curve. The model therefore nominally results in two meteorites of 3.4 g and three of 1.5 g. However, some minor flares, which were not modeled, occurred toward the end of the light curve. Some of the fragments were probably subject of additional mass loss in form of dust. Nevertheless, the dynamic measurements demonstrate that at least one fragment survived.

In summary, apart from relatively insignificant (in terms of lost mass) dust releases, only one severe fragmentation occurred.
It was at a height of 58 km under the dynamic pressure of 0.4~MPa. The meteoroid probably disrupted there into five pieces of similar masses. The relatively low mass of these fragments resulted into their large deceleration. The dynamic pressure did not increase so dramatically as would be the case for larger bodies and reached a maximum of 1.9~MPa at the height of 40~km (Fig.~\ref{pressure}). The velocity already decreased here to 24~km~s$^{-1}$. The strength of the material was large enough to withstand this dynamic pressure without further fragmentation. At the same time, the mass of the fragment was large enough for a small part of it to remain intact when the ablation ceased and to likely land as a meteorite.

\section{Discussion}

\subsection{The uniqueness of the Geminid EN131212\_041259}

The results of a detailed analysis of the observation of the bright Geminid EN131212\_041259 presented above led to one significant conclusion, namely that under certain conditions, which, however, can occur only very rarely, even such fast meteoroids can cause meteorites to fall. This alone proves its uniqueness. Firstly, because it is possible with the Geminids, as there is no other such case reliably documented in the literature, and on the other hand, because it is generally possible for a meteoroid with such a high entry velocity. 

\begin{figure*}[t]
    \centering
    \includegraphics[width=\hsize]{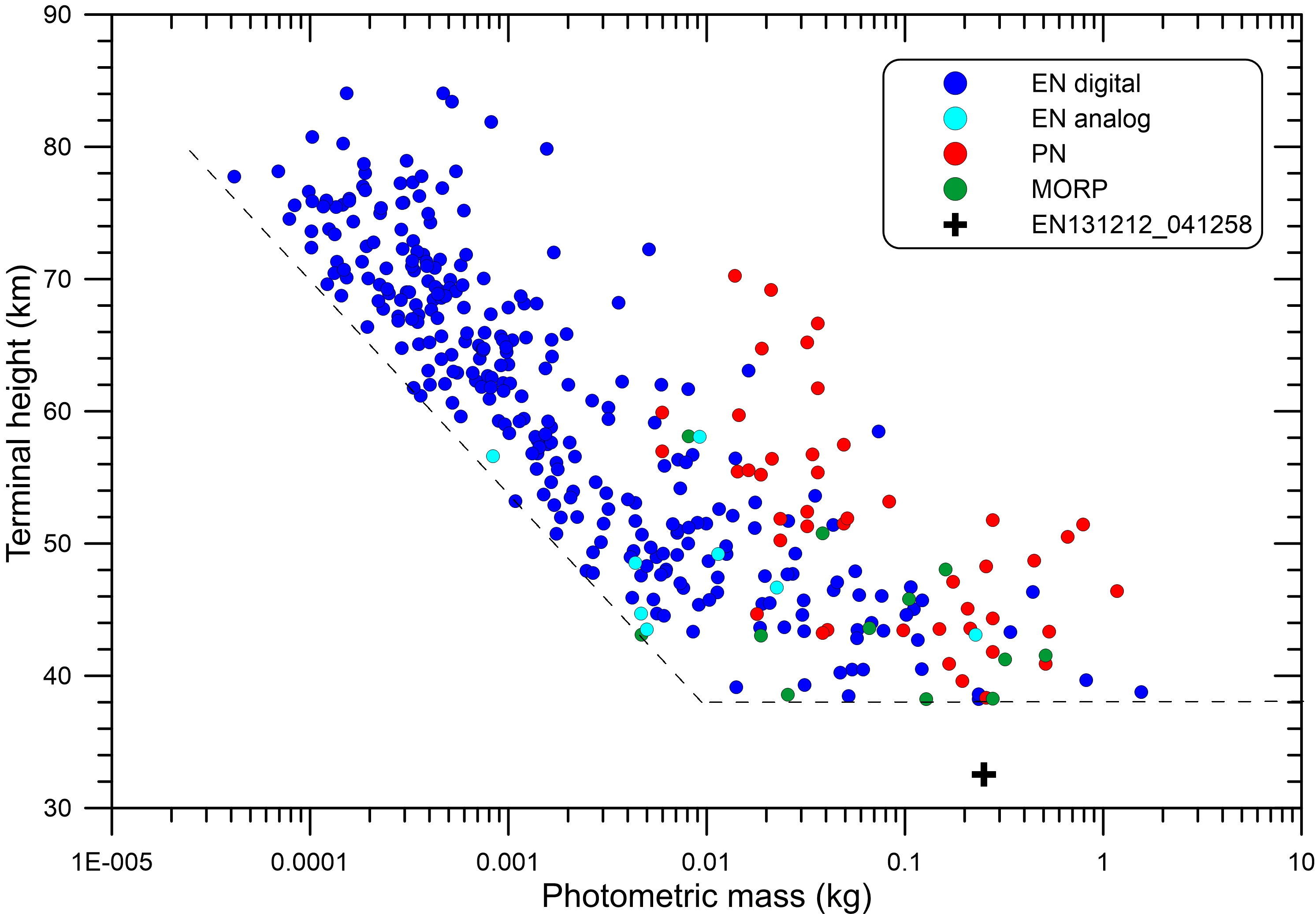}
    \caption{Dependence of the end heights of Geminid meteors and fireballs on their initial mass}
    \label{hemass}
\end{figure*}

It is therefore clear that this Geminid is unique in terms of the possibility of a meteorite fall, but is it also exceptional in other aspects compared to other Geminids?  From our own observations of bright Geminids in scope of the European Fireball Network (EN), as well as from published data from some other fireball networks, such as the Prairie Fireball Network (PN) in the USA \citep{mcc78} or the Canadian fireball network called Meteorite Observation and Recovery Project (MORP) \citep{hal78}, we know that it is definitely not particularly unusual in terms of how bright and massive this Geminid was. We chose data from these old bolide networks because one of the authors had used them earlier in his dissertation \citep{spu91}, so they were immediately available and contained the necessary data for really bright Geminids. And in all of these sets of data are Geminids which are even brighter and larger. So this Geminid is certainly not especial in terms of either absolute brightness or mass, which is a more meaningful parameter, because greater brightness can be caused by a single short flare during which a large amount of dust is released, but the entry mass may not be that high. It can be seen among others in Figure~\ref{hemass}.  However, uniqueness of the presented Geminid becomes apparent when we focus on the depth of penetration, i.e., its ablation resistance represented by a terminal height as is shown in the same Figure~\ref{hemass}. 

For this purpose, it is appropriate to select not only the largest Geminids, but also the widest possible range of masses, from which it should be possible to determine how Geminids generally behave in terms of penetration depth, i.e., terminal heights. This can then be compared with the Geminid mentioned in this article to determine how rare it is. Therefore, we used the largest representative sample of Geminids from our own data recorded by the European Fireball Network especially during its digital era, i.e. since 2014 \citep{Spu17,bor22}. To this, we added data on most of the Geminids that we recorded in our network during the years 2014--2025.  For some years, the data are complete (2014-2016, 2020-2024), while for other years, the data are complete only for brighter cases (brighter than magnitude $-8$) and for fainter cases, it is mostly only for longer and well defined Geminids. In total, there are about 290 Geminids in this set of digitally recorded fireballs. Along with this data, we also used older EN data from analog cameras, but there are not many such cases, only 8 in total. Therefore, we used also above mentioned 44 bright PN fireballs from \cite{mcc79} and 12 MORP fireballs from \cite{hal88} to these data sets. Because the terminal altitude is only slightly dependent on the sensitivity of the technique used due to the rapid drop in brightness at the end, therefore, it does not matter so much that we are comparing end heights obtained with techniques of different sensitivity. However, in order to use all the data together, it was necessary to have as comparable a mass scale as possible, i.e., for the photometric masses to be based on the same luminous efficiency. Therefore, the sets of the original masses from PN and MORP Geminids were recalculated with luminous efficiency according to the work \cite{rev01}, which we use for the analysis of bolides from the EN as explained in \cite{bor22}. The resulting graph of the dependence of the terminal height of  Geminids from the above mentioned sources on the initial mass is shown in Figure~\ref{hemass}.

This graph clearly shows that the Geminid we present is also unique in terms of penetration depth compared to a large number of others. No other Geminid penetrated to a height lower than 38 km. It should also be mentioned that our Geminid was recorded using the technique with the lowest sensitivity of all. In this context, it is important to note that the current instruments used systematically in EN since 2014, i.e., Digital Autonomous Fireball Observatory (DAFO) as described in \cite{Spu17} in combination with even more sensitive IP video cameras, clearly have the highest sensitivity of all the observation systems used. Of course, this depends on the observation conditions, the distance of the bolide from the instruments, and other circumstances, but on average, this means a higher sensitivity of about 2-3 magnitudes. Similarly, photographic records from the PN and MORP networks had higher sensitivity than the all-sky EN images, although not by much. It follows, therefore, that the terminal altitude value of 33.53 km is rather disadvantaged compared to records from other systems, and at the very least, it cannot be argued that such a low terminal altitude is due to suitable observation conditions or suitably located stations. As can be seen from the Table~\ref{camerastable}, the terminal altitude from the second station, Chur\'{a}\v{n}ov, is still significantly below the specified limit of 38 km from the distance of 125 km. 

This graph also clearly shows how rare such cases are. Neither during the entire period of meteor and fireball monitoring within the European Fireball Network and the previous two-station meteor observation program led by Z. Ceplecha at the Astronomical Institute of the Czech Academy of Sciences in Ond\v{r}ejov since 1951, nor during the existence of the Prairie Fireball Network in the USA and the Canadian MORP Fireball Network, has any other Geminid penetrated the atmosphere so deeply. However, this graph showed another, even more interesting result regarding the physical properties of Geminids. Specifically, for small Geminids, the usual dependence of the final altitude on the increasing entry mass of the meteoroid applies, which is not surprising. But it is interesting how sharply this area is defined, which means that among small Geminids, i.e., those weighing approximately 10 grams or less, there is no material stronger than the limit given by the dashed line. However, at a mass of approximately 10 grams and an altitude of 38 km, there is a significant change, where this original dependence suddenly ends and it no longer matters how massive the original meteoroid was, and the terminal altitude limit remains constant. This applies to at least two orders of magnitude of Geminid masses up to a mass of approximately 1 kg. This indicates that, with very rare exceptions, such as the Geminid presented in this article, there is no material among the Geminids that is stronger and more resistant than this limit. We can therefore generalize that the usual strength of Geminids is in the vast majority of cases less than or at most equal to that given by this limit of terminal altitude around 38 km. 

The strength of the original meteoroid is often associated with the maximum dynamic pressure to which the meteoroid is exposed. As shown in Section 3.4, the Geminid presented here reached a maximum pressure of 1.9~MPa at an altitude of 40 km. Figure~\ref{hemass} shows that a total of 11 other Geminids (6 EN, 3 MORP, and 2 PN) reached the 38 km limit within a range of 2 km, i.e., between the terminal altitudes of 40 and 38 km. With one exception, the most massive Geminid weighing 1.7~kg \citep[EN121218\_193710 described in the work of] []{hen}, when the maximum dynamic pressure reached only 0.88 MPa at an altitude of 46.3 km, all the others reached a maximum pressure in the range of 1.2--1.8 MPa. For all of them, this was in the range of altitudes between 41.6 and 44.3 km.  It is therefore clear that the material of these Geminids is very similar to each other, and according to the work of ~\cite{2strengths}, such high strength corresponds to the category to which, among others, confirmed meteorite falls belong. However, in observed meteorite falls, these pressures can reach even around 10~MPa for the strongest meteoroids, and these are usually ordinary chondrites. From this, we can conclude that the Geminid material is the strongest of all other well-described meteor showers and that, in very rare cases, its strength corresponds to that of some meteoroids belonging to observed meteorite falls. A very significant limiting factor for meteorite falls originating from the Geminids is therefore not so much the structure of the Geminid material, but primarily their high collision speed. Only under favorable conditions can macroscopic material survive the atmospheric passage. If, in the present case, the fragmentation around the height 58 km did not occur, the dynamic pressure of 1.8 MPa would be reached at a height of 44 km with the speed still exceeding 30 km s$^{-1}$. The fragmentation would probably occur there. But all produced fragments larger than about 10 g would reach 2 MPa shortly afterwards followed probably by their further destruction. Fragments smaller than 10 g could avoid further
fragmentation but due to continued ablation they would not produce meteorites larger than one gram. The early fragmentation followed by increased deceleration was therefore crucial for dropping a meteorite (small but non-negligible).

From the values of the maximum dynamic pressure of the strongest Geminids, it is also clear that Geminids do not belong to the strongest meteoroids producing observed meteorite falls, which include ordinary chondrites. This may imply that the most likely known material corresponding to the Geminids could be some type of carbonaceous chondrites. This corresponds to the conclusions of the article by \cite{hen24}, in which several EN Geminids were modeled in detail and found to have similarly high mechanical strength. As a result, a material corresponding to carbonaceous chondrites was also proposed here.     

\subsection{Revelation of unsubstantiated claim on the same topic}

\begin{figure}
    \centering
    \includegraphics[width=\linewidth]{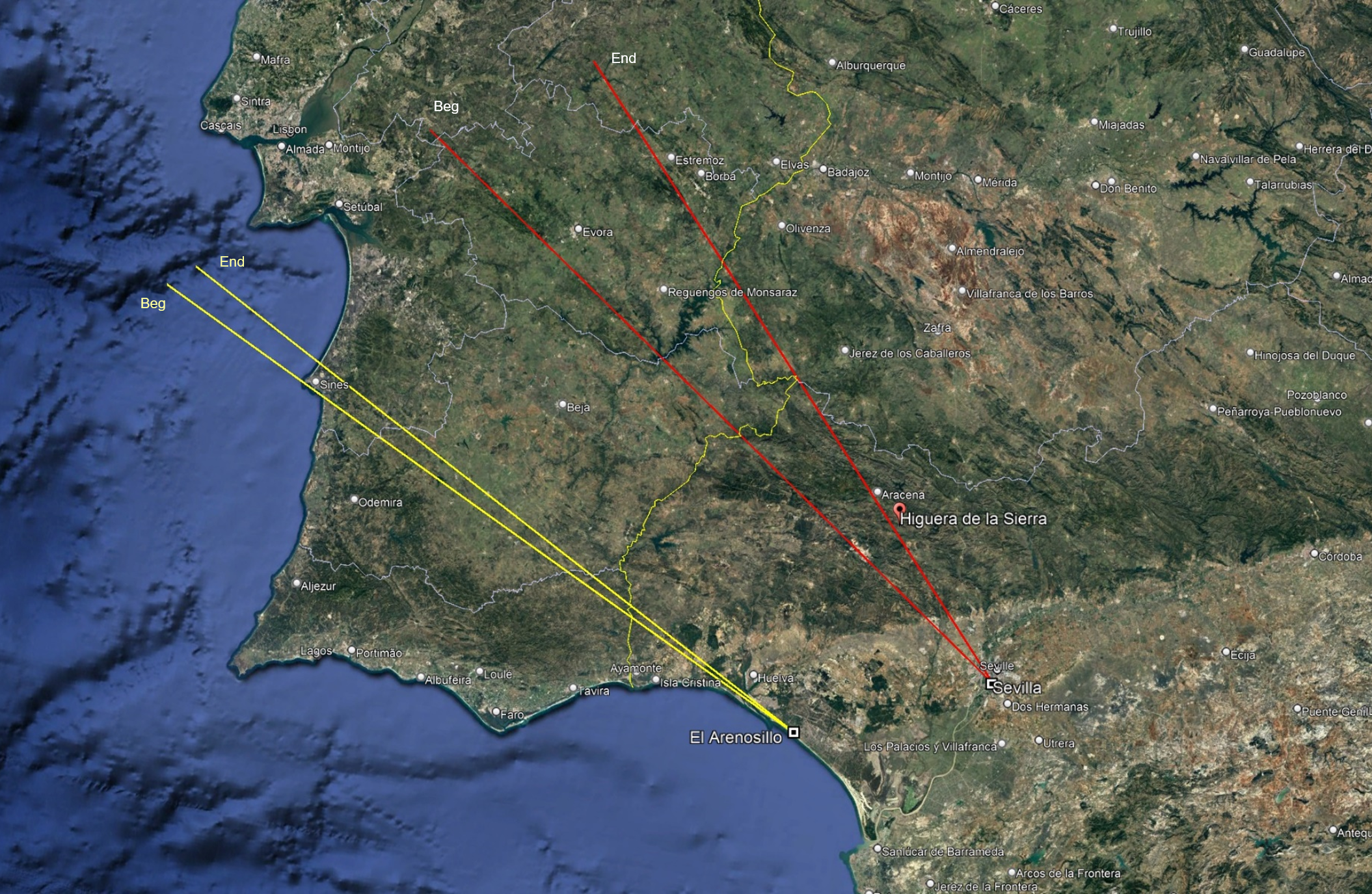}
    \caption{Azimuth range of the SPMN Geminids as recorded from stations Sevilla (red)) and El Arenosillo (yellow); source of the map: Google Earth}
    \label{SpGem}
\end{figure}

A preliminary account of our observation was given at the Meteoroids 2013 conference\footnote{https://www.astro.amu.edu.pl/Meteoroids2013/main\_content/data/ abstracts.pdf}. The main reason why we did not publish it immediately after is that an even more remarkable Geminid, with a terminal height as low as 24.8 km, recorded by the Spanish Meteor Network (SPMN), was claimed by \cite{mad13} shortly after the conference. However, as we now explain, our more detailed assessment of those data demonstrates this analysis was flawed. Some details in this article did not seem right to us from very beginning such as the relatively low initial altitude and an extremely low terminal altitude, which looked as if the altitude scale had been shifted significantly towards lower altitudes by at least 10 km. \cite{mad13} indicate that the fireball was observed from two Spanish stations, Sevilla and El Arenosillo, whose geographical coordinates, and the timing of the event, are given in the paper. Low-resolution images of the fireball, and its apparent trajectory in the sky as observed from the two stations were also provided (their Fig. 1). Besides this, key information was missing, such as coordinates at any point of the trajectory, a map of the trajectory, or information on the beginning and terminal altitudes separately for each station. Nonetheless, based on the provided timing and coordinate information, our own reconstruction of the trajectory on the sky matched well that shown in \cite{mad13}. However, upon calculating the azimuth and elevation above the horizon for the beginning and end of the fireball (Table~\ref{tabspmn}) and plotting the azimuth on a geographical map (Fig.~\ref{SpGem}), we discovered that what the authors referred to as a single bolide in the images was in fact two different bolides, both of which likely belonging to the Geminids. The observed directions from the two stations do not even intersect. As it is obviously impossible to calculate the trajectory of a bolide from directions that do not intersect, the claim by \cite{mad13} is entirely spurious.

\begin{table}
\caption{Azimuth and elevation for Sevilla and El Arenosillo stations for the SPMN Geminid from December 15, 2009 at $4^{\rm h}20^{\rm m}11\fs9$~ UT}
\label{tabspmn}
\centering
\begin{tabular}{lcccc}
\hline\hline
 & \multicolumn{2}{c}{Sevilla} & \multicolumn{2}{c}{El Arenosillo} \\
\cline{2-3}\cline{4-5}
\noalign{\smallskip}
 & Az ($^\circ$, N=0) & H ($^\circ$) & Az ($^\circ$, N=0) & H ($^\circ$) \\
 \noalign{\smallskip}
\hline
\noalign{\smallskip}
Beg & 306.53 & 43.86 & 297.03 & 17.34 \\
End & 319.32 & 19.59 & 299.36 & 10.02 \\
\hline
\end{tabular}
\end{table}

\section{Conclusions}

The main purpose of this work was to publish observations and a detailed analysis of one very exceptional Geminid fireball designated EN131212\_041259, which was recorded in the Czech part of the European Fireball Network on December 13, 2012. At the same time, we explained why we are publishing this result with such a delay. On the other hand, thanks to the 13-year time gap, the significant improvement in the efficiency of observations within the EN has made it possible to accumulate a large number of other Geminids, which, together with older data from some fireball networks (PN, MORP), clearly demonstrate the uniqueness of this single Geminid. 

\setlength{\parindent}{0pt} The main conclusions of this work are therefore as follows: 
\setlength{\parskip}{1em}

\setlength{\parindent}{10pt}
The analyzed Geminid fireball, recorded by the European Fireball Network, provides compelling observational evidence that meteorite survival from the Geminid stream, although extremely rare, cannot be excluded. Despite an initial entry velocity of 35.75~km~s$^{-1}$, the fireball exhibited an unusually deep atmospheric penetration, reaching a terminal height of 32.5~km with a measured terminal velocity of 6.8~km~s$^{-1}$. Such end-of-luminous-flight parameters are comparable to those observed in confirmed meteorite-producing fireballs at substantially lower entry velocities. Fragmentation modeling indicates the possible survival of at least one meteorite of mass 3 -- 5 grams. When interpreted in the context of established fragmentation theory and empirical constraints derived from rare high-velocity meteorite falls, this event represents the strongest evidence to date that compact and mechanically strong Geminid fragments may under favorable circumstances survive atmospheric passage. The potential recovery of Geminid material would offer a unique opportunity to directly investigate the physical and chemical properties of meteoroids associated with asteroid (3200)~Phaethon. It, nevertheless, follows from our analysis that only gram-sized meteorites can be expected from Geminids since larger fragments would be exposed to dynamic pressures they cannot withstand.

\setlength{\parskip}{0em}

Given the very high impact speed of almost 36 km~s$^{-1}$, we can generalize this conclusion to other fireballs, i.e., under certain very special conditions, even such a fast fireball can end up as a meteorite. 

From the data available to us, we have shown that this Geminid penetrated significantly deeper into the atmosphere than all Geminids recorded in the EN, PN, and MORP fireball networks. And we are not aware of any other observation programs that have recorded a comparable Geminid. 

From the dependence of the terminal heights of Geminids on the entry mass in the range of 5 orders of magnitude, we found that for low-mass Geminids from one tenth of a gram to approximately ten grams, there is a significant dependence on the terminal height, i.e., heavier meteoroids penetrate deeper into the atmosphere, which was to be expected. However, the behavior of bright Geminids is surprising, as for masses from ten grams to one kilogram, the penetration depth does not go below 38~km, which is a barrier for all Geminids observed so far, with the exception of the one described in this paper. Its terminal altitude was a record 32.5~km. We have demonstrated that the terminal height of 24.8 km reported
by \citet{mad13} for a Geminid observed in Spain is not trustworthy at all. 

At the same time, our Geminid proved to be the strongest of all recorded, reaching a maximum dynamic pressure of 1.9~MPa. However, even among them there are very strong meteoroids, i.e., those that are close to the 38~km barrier in terms of penetration depth and have maximum dynamic pressures in the range of 1.2-–1.8~MPa, which correspond, among other things, to the pressures of most of observed meteorite falls \citep{2strengths}. Our investigation confirmed the conclusion of \citet{hen}, who modeled selected Geminids in detail, that medium sized Geminids (20--200 g) are the most compact ones.

A significant feature of Geminid EN131212\_041259 is the observed regular fluctuations in the first third of the light curve with a period showing a marked accelerating tendency in the range of 20-70~Hz. As we have found from our observations, this is a relatively common phenomenon observed not only in longer Geminid fireballs.  This is known as flickering, which is described for Geminids, for example, in \cite{hal88}. In contrast, in the second part of flight on the light curve, we observe very numerous millisecond flashes similar to sparking. This was first described in \cite{spu08}. Neither of these phenomena has yet been satisfactorily explained.  

From all the Geminids observed by the EN, PN, and MORP fireball networks, it is clear that the Geminid meteor shower does not contain bodies that are significantly heavier than 1~kg. This contrasts with the Taurid meteor shower, where there are proven Taurids with a mass at least three orders of magnitude higher, as published, for example, in \cite{Spu17}.

Similarly, a comparison of all Geminids used in Figure~\ref{hemass} shows that, in addition to being very resistant and solid meteoroids, they also contain relatively fragile material corresponding to type II bolides according to \cite{Ceplecha1976}. It is therefore clear that, in terms of physical and structural properties, Geminids are relatively heterogeneous meteoroids.

As for the used and previously unpublished data for EN Geminids, we only used the values of the end heights and entry masses in this work. Of course, the data is much more complex and includes information on atmospheric trajectories, physical properties, and heliocentric orbits. We will deal with all of this in another article, where we will, of course, publish all this data. As already mentioned, this article is mainly devoted to that one extraordinary Geminid, which is also described in detail here.

\begin{acknowledgements}
The authors acknowledge the work of all operators and technicians who are keeping the observatories of the European
Fireball Network and the associated infrastructure running. Special thanks go to L. Kotkov\'{a} for data preparation and measurement, J. Kecl\'{\i}kov\'{a}, P. V\'{a}chov\'{a}, H. Zichov\'{a}, and M. Macourkov\'{a} for measurement of all-sky images and L. Kop\v{r}ivov\'{a}, L. Shrben\'{y} and R. \v{S}tork for measurement of records from IP cameras. Our special thanks go also to J. Svore\v{n} and D. Tomko (Astronomical Institute of the Slovak Academy of Sciences) and H. Mucke and T. Weiland (Österreichische Astronomische Verein) for their support of the network in Slovakia and Austria, respectively.
We would also like to thank the anonymous reviewer for his valuable comments and careful editing of the text.
This work was supported by grant 24-10143S from the Czech Science Foundation (GA ČR). The recent modernization
of the network, crucial for this work, was funded from the Praemium Academiae of the Czech Academy of Sciences. The institutional research plan is RVO:67985815.
\end{acknowledgements}

\end{document}